\documentclass[onecolumn,showpacs,preprintnumbers,amsmath,amssymb,unsortedaddress]{revtex4} 

\usepackage[utf8]{inputenc}  
\usepackage{graphicx}        
\usepackage{dcolumn}         
\usepackage{bm}              
\usepackage{subfigure} 
\usepackage{ltablex}
\usepackage{xcolor}

\graphicspath{{Figures/},.}         

\begin{document}


\title{Persistent collective trend in stock markets}

\author{Emeric Balogh}
\affiliation{Babeş-Bolyai University, Department of Theoretical
  Physics, RO-400084 Cluj-Napoca, Romania}

\author{Ingve Simonsen} 
\email{Ingve.Simonsen@phys.ntnu.no}
\homepage{http://web.phys.ntnu.no/~ingves} 
\affiliation{Department of
  Physics, Norwegian University of Science and Technology (NTNU),
  NO-7491 Trondheim, Norway}

\author{B\'alint \ Zs.\ Nagy} 
\affiliation{Babeş-Bolyai University, Faculty of Economics and
  Business Administration RO-400591 Cluj-Napoca, Romania}

\author{Zolt\'an N\'eda}
\affiliation{Babeş-Bolyai University, Department of Theoretical
  Physics, RO-400084 Cluj-Napoca, Romania}

\date{\today}



\begin{abstract}
  Empirical evidence is given for a significant difference in the
  collective trend of the share prices during the stock index rising
  and falling periods. Data on the Dow Jones Industrial Average and
  its stock components are studied between 1991 and 2008. Pearson-type
  correlations are computed between the stocks and averaged over
  stock-pairs and time. The results indicate a general trend: whenever
  the stock index is falling the stock prices are changing in a more
  correlated manner than in case the stock index is ascending. A
  thorough statistical analysis of the data shows that the observed
  difference is significant, suggesting a constant-fear factor among
  stockholders.
\end{abstract}

\maketitle


The world is once again experiencing a major financial-economic
crisis, the worst since the crash of Oct. 1929 that initiated the
great depression of the 1930s.  Many citizens are concerned for
obvious reasons; we are facing global recession; banks and financial
institutions go bankrupt; companies struggle to get credit and many
are forced to reduce their workforce or even go out of
business. Interest rates are increasing while private savings invested
in the stock market evaporate. Large parts of our contemporary
societies are deeply affected by the new financial reality.

The current financial crisis is one particular dramatic example of
collective effects in stock markets~\cite{1,2,3,Sornette}; during
crises nearly all stocks drop in value simultaneously. Fortunately,
such extreme situations are relatively rare. What is less known,
however, is that during more normal ``non-critical'' periods,
collective effects do still represent characteristics of stock markets
that in particular influence their short time behavior. One such
effect will be addressed in this publication, where our aim is to
present empirical evidence for an asymmetry in stock-stock
correlations conditioned by the size and direction of market moves.
In particular, we will present empirical results showing that when the
Dow Jones Industrial Average~(DJIA) index (``the market'') is
dropping, then there exists a significantly stronger stock-stock
correlation than during times of a raising market.  Our results
indicate that such enhanced (conditional) stock-stock correlations are
not only relevant during times of dramatic market crashes, but instead
represents features of markets during more ``normal'' periods.

Distribution of returns is traditionally used as one of the proxies
for the performance of stocks and markets over a certain time
history~\cite{1,2,3}.  In the economics, finance and econometrics
literature the problem of market sentiment and investor confidence is
usually addressed by the use of various indicators. These indicators
are either derived from objective market data~\cite{Bodie}, or
obtained by conducting questionnaire-based surveys among professional
and individual investors~\cite{Shiller}. In the present study we
consider thus the first approach, since we believe that the market
data (prices and returns) are more objective proxies than
questionnaire-inferred data.

The basic quantity of interest is the logarithmic return, defined as
the (natural) logarithm of the relative price change over a fixed time
interval $\Delta t$, {\it i.e.}:
\begin{align}
  r_{\Delta t}(t) &= \ln \left( \frac{p(t+\Delta t)}{p(t)} \right),
  \label{eq:1}
\end{align}
where $p(t)$ denotes the asset price at time $t$~\cite{1,2,3}. In
addition to this basic quantity, it is also desirable to have
available a time-dependent proxy where the asset performance is gauged
over a non-constant time interval. One such approach is the so-called
{\em inverse statistics approach}~\cite{4,5,6,7} recently introduced
and adapted to finance from the study of turbulence~\cite{8,8-2}.  The
main idea underlying this method is to {\em not} fix the time interval
(or window), $\Delta t$ in Eq.~(\ref{eq:1}), but instead to turn the
question around and ask for what is the (shortest) waiting time,
$\tau_\rho=t-t_0$, needed to reach a given (fixed) return level,
$\rho$, for the first time when the initial investment was made at
time $t_0$ (see Ref.~\cite{4} for details). Hence, the inverse
statistics approach concerns itself with the study of the distribution
of waiting times~\cite{9} that in the following will be denoted by
$p(\tau_\rho)$.

\begin{figure}[t]
  \begin{center}
  \subfigure {\includegraphics[width=0.35\columnwidth]{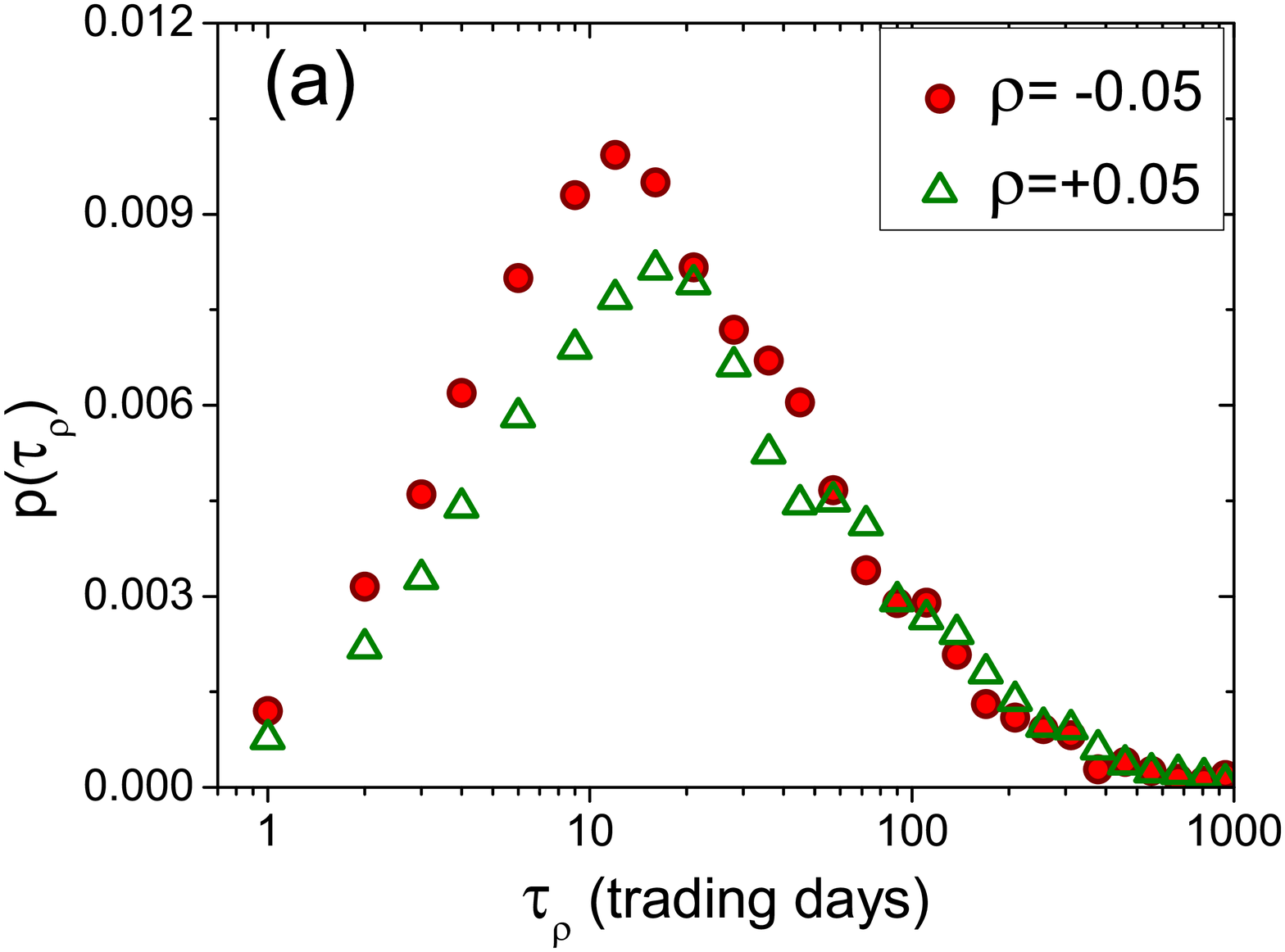}}
  \subfigure {\includegraphics[width=0.35\columnwidth]{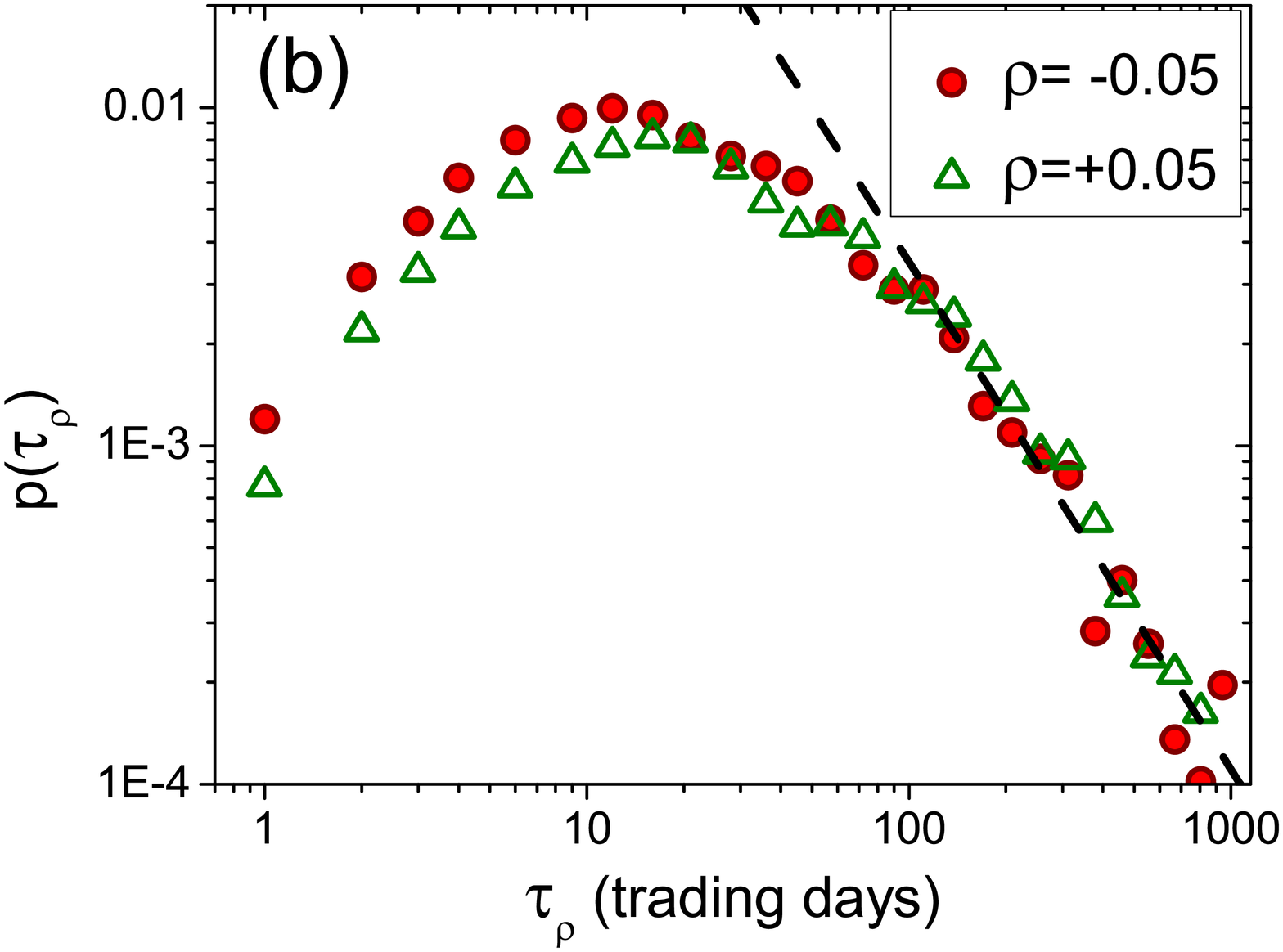}}
  \end{center}
   \caption{Inverse statistics results for logarithmic return levels
     of $\rho=\pm 5\%$ for the DJIA index (data between $1991$ and
     $2008$). The figures show the gain-loss asymmetry; open green
     triangles represents $\rho>0$, while filled red circles refer to
     $\rho<0$.  On the log-linear scale (a) the asymmetry is more
     evident, while on log-log scale (b) the power-law nature of the
     tail of the distribution is observable. The dashed line indicates
     the slope $-3/2$.}
  \label{Fig:1}
\end{figure}

Recently, this method of analysis has been applied to the study of
various single stocks and market indices, both from mature and
emerging markets, as well as to foreign exchange data and even
artificial
markets~\cite{4,5,6,7,10,12,12A,Poland,Austria_Poland,Korea,Poland-2,Saxo-2}. The
waiting time histograms possess well-defined and pronounced
($\rho$-dependent) maxima~\cite{4}
followed by long power law tails, $p(\tau_\rho)\sim \rho^{-\alpha}$,
with $\alpha\simeq 3/2$~(Figs.~\ref{Fig:1}) a value that is a
consequence of the uncorrelated increments of the underlying asset
price process~\cite{9}.

\begin{figure}[t]
  \begin{center} 
  \subfigure {\includegraphics[width=0.35\columnwidth]{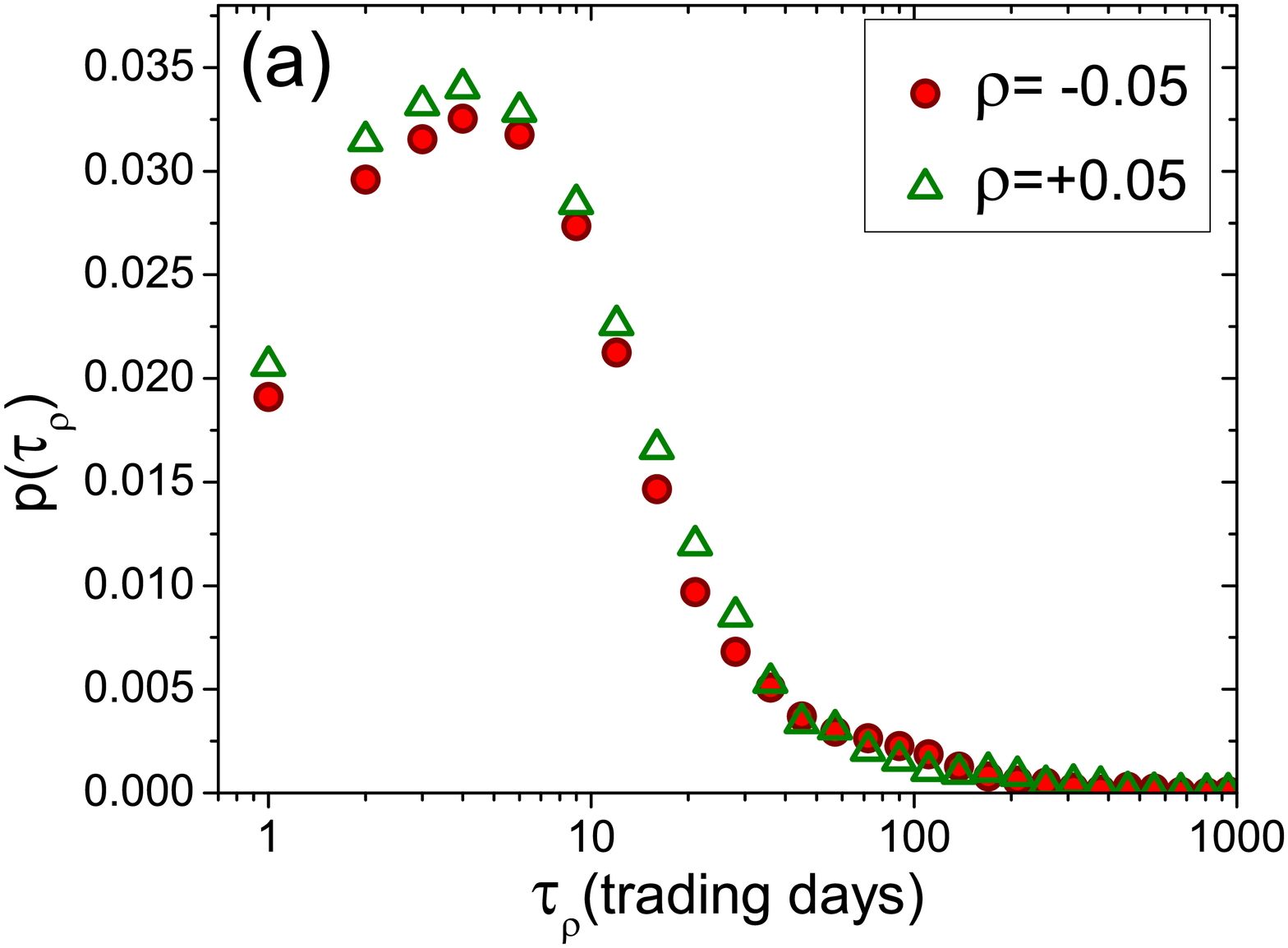}}
  \subfigure {\includegraphics[width=0.35\columnwidth]{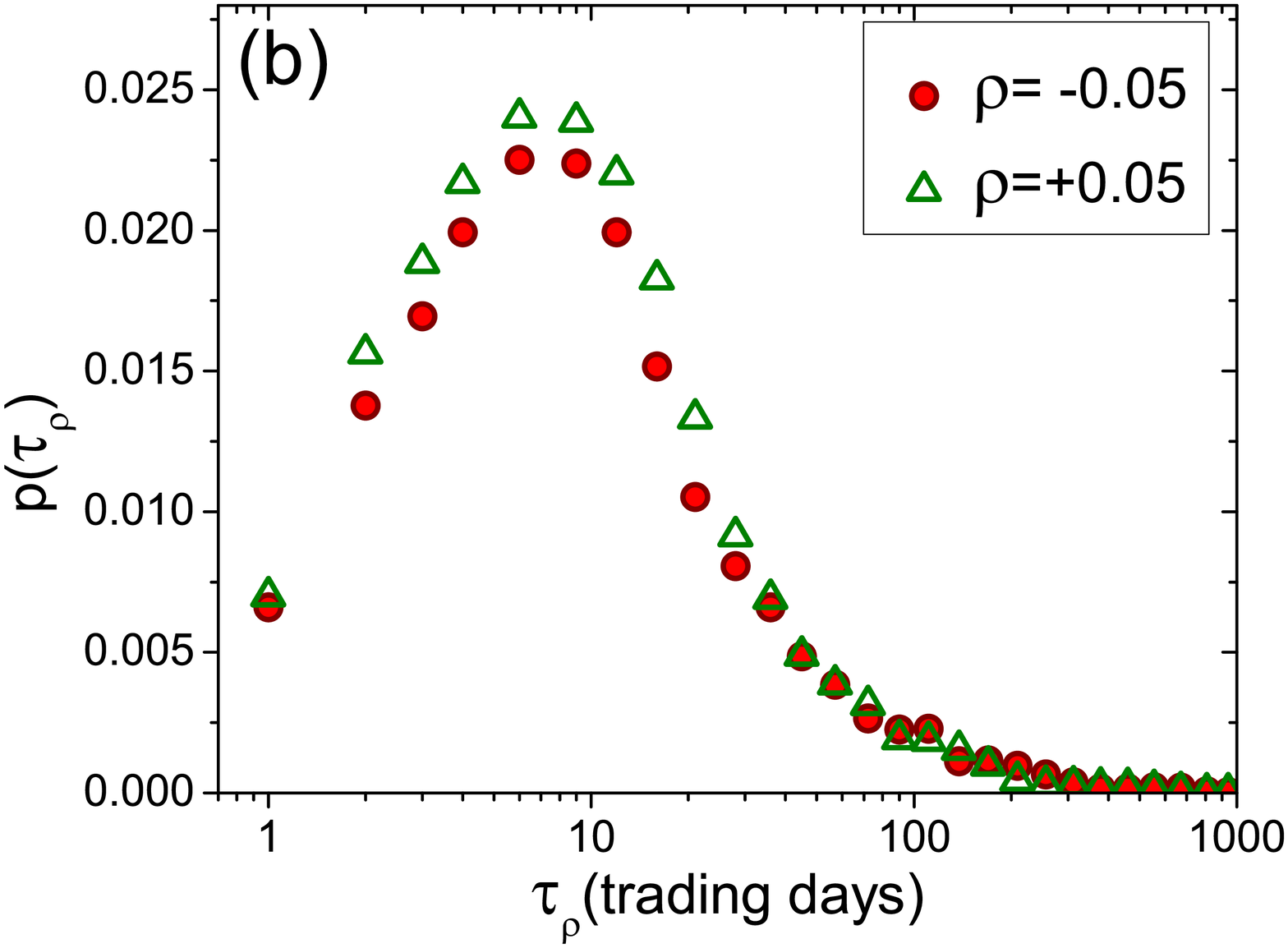}}
  \end{center}
  \caption{Same as Fig.~\ref{Fig:1}(a) but now for the stocks (a)
    General Motors and (b) McDonald’s Corp., both being part of the
    DJIA index.  Notice that gain-loss asymmetry is not observed in
    this case.  }
  \label{Fig:2}
\end{figure}

Studies of single stocks, for given (moderate) positive and negative
levels of returns, $\pm|\rho|$, have revealed, almost symmetric
waiting time distributions
(Figs.~\ref{Fig:2})~\cite{12,Saxo-2,13}. Unexpectedly, however, stock
index data seem not to share this feature. They do instead give raise
to asymmetric waiting time distributions (Fig.~\ref{Fig:1}a) for
return levels $|\rho|$ for which the corresponding single stock
distributions were symmetric~\cite{12,13}.  This asymmetry is
expressed by negative return levels being reached {\em sooner} than
those corresponding to positive levels (of the same magnitude of
$\rho$).  This effect was termed the {\em gain-loss
  asymmetry}~\cite{4} and has later been observed for many major stock
indices~\cite{4,5,Our_Unpublished,Austria_Poland,Korea,Saxo-1}. It is
here important to note that the gain-loss asymmetry is not a
consequence of the generally long-term positive trend (or drift) of
the data since this was removed by considering an average with a
suitable window size on the prices.  The long-term positive trend will
affect long waiting times and would induce shorter waiting times for
the positive return levels.  However, empirically one finds that the
waiting times of indices are shortest for negative return levels ---
the opposite of what is to be expected from the long term trend
effect. In passing we note that recently it has been found that also
single stocks may show some degree of gain-loss asymmetry when the
level of return, $|\rho|$, is getting sufficiently
large~\cite{Saxo-2,Peter}. However, it still remains true that for not
too large return levels, {\it e.g.}  $|\rho|=0.05$, the waiting time
distributions for single stocks are symmetric to a good
approximation~\cite{Saxo-2}.

The presence of this asymmetry may seem like a paradox since the value
of a stock index is essentially the (weighed) average of the
individual constituting stocks. Even so, one does observe an
asymmetric waiting time distribution for the index comprised of
(more-or-less) symmetric single stocks. How can this be rationalized?
Recently, a minimal (toy) model --- termed the {\em fear factor model}
--- was constructed for the purpose of explaining this apparent
paradox~\cite{13}. The key ingredient of this model is the so-called
collective fear-factor, a concept similar to
synchronization~\cite{14}.  At certain times, controlled by a
``fear-factor'', the stocks of the model {\em all} move downwards,
while at other times they move independently of each other. This is
done in a way that the price processes of the single stocks are (over
a long time period) guaranteed to produce symmetric waiting time
distributions (and uncorrelated price increments). The fear-factor
model, that qualitatively reproduces well empirical findings,
introduces collective downward movements among the constituting
stocks.  The model synchronizes downward stock moves, or in other
words, it has stronger stock-stock correlations during dropping
markets than during market raises. This means that the fear-factor of
the stockholders is stronger than their optimism-factor on
average. This is consistent with the findings of Kahneman and
Tversky~\cite{15}, reported in the economics literature, that
demonstrate that the utility loss of negative returns is larger than
the utility gain for positive returns in the case of most investors.

Recently, the idea of the fear-factor model~\cite{13} was reconsidered
and generalized by Siven {\it et al.}~\cite{Saxo-1} by allowing for
longer time periods of stock co-movement (correlations). These authors
also find that the gain-loss asymmetry is a long timescale
phenomena~\cite{Saxo-1}, and that it is related to some correlation
properties present in the time series~\cite{Saxo-2}.  It was also
proposed that the gain-loss asymmetry is in close relationship with
the asymmetric volatility models (E-GARCH) used by
econometricians~\cite{Nelson}.

Furthermore, also additional explanations for the gain-loss asymmetry
have been proposed in the literature. Those include the leverage
effect~\cite{2,Leverage,Peter_Leverage,Saxo-3}, and regime switching
models~\cite{Peter}. So far, it is fair to say that the cause of the
gain-loss asymmetry is still partly debated in the literature.

The key idea of the fear factor model~\cite{13,Saxo-2} 
is the enhanced stock-stock correlations during periods of falling
  market. Up to now this idea has not been supported by empirical data.
In this Letter, we conduct such a delicate statistical analysis, and
we are able to show, based on empirical data, that indeed there
exist a stronger stock-stock correlations during falling as compared
to rising market. 

\bigskip

Let $r_{\Delta t}^{\,x}(t)$ denote the logarithmic return of stock $x$
(from the index under study) between time $t$ and $t+\Delta t$ (the
time unit in the DJIA data is one trading day).  In order to
facilitate the coming discussion, we introduce the following notation
for an arithmetic average taken over a set, say ${\mathcal A} =
\left\{A(t)\right\}_{t=t_1}^{t_2}$:
\begin{align}
  \left< \mathcal A \right>
  = \left< \left\{A(t)\right\}_{t=t_1}^{t_2} \right>  
  = \frac{ \displaystyle\sum_{t=t_1}^{t_2}  A(t)}{\left|{\mathcal A}\right|} 
  = \frac{\displaystyle \sum_{t=t_1}^{t_2} A(t)}{t_2-t_1+1},
      \label{eq:average}
\end{align}
where $\left|{\mathcal A}\right|$ denotes the cardinality of the set,
{\it i.e.} the number of elements in ${\mathcal A}$. If no explicit
limits are given for the set (like in $\{ A(t)\}_t$), all possible
values will be assumed. In terms of this notation, a Pearson-type
correlation can then be computed between each stock pair $(x,y)$
resulting in the following (equal time) {\em stock-stock correlation
  function}
\begin{align}
  S_{(x,y)}(t,\delta t,\Delta t) 
    &=  
    \frac{ 
      \Big<
          \big\{ r^{\,x}_{\Delta t}(t') \, r^{\,y}_{\Delta t}(t') \big\}_{t'=t}^{t+\delta t} 
       \Big>
       - \Big< \big\{ r^{\,x}_{\Delta t}(t')\big\}_{t'=t}^{t+\delta t} \Big> 
         \Big< \big\{ r^{\,y}_{\Delta t}(t')\big\}_{t'=t}^{t+\delta t} \Big> 
   }{ 
     \sigma_{\Delta t}^{\,x}(t;\delta t)\, \sigma_{\Delta t}^{\,y}(t;\delta t) 
   }, 
  \label{eq:2}
\end{align}
where $\sigma_{\Delta t}^{\,\alpha}(t;\delta t)$ signifies the
volatility of stock $\alpha$ ($\alpha=x,y$) at time $t$ (and time
window $\delta t$), and is defined as
\begin{align}
  \sigma_{\Delta t}^{\,\alpha}(t;\delta t) &=
  \sqrt{ \Big< \big\{ \left[r^{\,\alpha}_{\Delta t}(t')\right]^2\big\}_{t'=t}^{t+\delta t}\Big> 
    - \Big< \big\{ r^{\,\alpha}_{\Delta t}(t')\big\}_{t'=t}^{t+\delta t}\Big>^2 }.
  \label{eq:3}
\end{align}
Note that $S_{(x,y)}(t,\delta t,\Delta t)$ contains two time scales;
$\delta t$ is the time window over which the average in
Eq.~(\ref{eq:2}) is calculated, while $\Delta t$ is the time interval
used to define returns ({\it cf.} Eq.~(\ref{eq:1})).

By definition, the stock-stock correlation function,
$S_{(x,y)}(t,\delta t,\Delta t)$, is specific to the asset pair
$(x,y)$, and does therefore not represent the market as a
whole. However, in order to obtain a representative level of
stock-stock correlation for the market (index) as a whole, we propose
to average $S_{(x,y)}(t,\delta t,\Delta t)$ over all possible stock
pairs $(x,y)$ contained in the index. In this way, we are lead to
introducing the {\em market component correlation function}
\begin{align}
  S_0(t,\delta t, \Delta t) &= 
  \left< \left\{ S_{(x,y)}(t,\delta t, \Delta t) \right\}_{\{(x,y)\}}\right>.
  \label{eq:4}
\end{align}
In passing, we note that the average contained in Eq.~(\ref{eq:4})
potentially should be weighted so that the contribution to the
correlation function $S_0(t,\delta t, \Delta t)$ from a stock pair
$(x,y)$ is weighted with a factor that is proportional to the sum of
the weights associated with the two stocks and used to construct the
value of the index. Typically this weight corresponds to the
capitalization of the company in question. Since we here, however, are
studying the DJIA --- for which all constituting stocks have the same
weight in the index (an atypical situation) --- this possibility has
not been considered here and neither has the weight factor been
included in the definition of $S_0(t,\delta t, \Delta t)$.

The market component correlation function, as defined by
Eq.~(\ref{eq:4}), measures the overall level of stock-stock
correlations of the index (market) under investigation {\em
  independent} of the market is raising or falling. However, what we
have set out to study, is if there exists any significant difference
between these two cases. To this end, we introduce what we below will
refer to as the {\em conditional market component correlation
  function}, $C_0(\rho,\delta t, \Delta t)$, that measures the typical
value of the market component correlations $S_0(t,\delta t, \Delta t)$
{\em given} that the (logarithmic) return of the index itself,
$r_{\delta t}(t)$ is above (below) a given return threshold value
$\rho$. Mathematically, the conditional market component correlation
function is defined by the following conditional time average
\begin{subequations}
  \label{eq:5}
\begin{align}
  C_0(\rho,\delta t, \Delta t) &=
  \Big< \big\{{\mathbb  C}(\rho, t,\delta t, \Delta t) \big\}_t \Big>,
  \label{eq:5a}
\end{align}
where a  time-dependent conditional market component correlation
function has been introduced as:
\begin{align}
  \label{eq:5b}
  \{{\mathbb C}(\rho,t,\delta t,\Delta t)\}_t   &=
  \begin{cases}
    \big\{ S_0(t,\delta t, \Delta t)\, | \,r_{\delta t}(t)\geq \rho \big\}_t 
       & \mbox{if } \rho\geq 0 \\
    \big\{ S_0(t,\delta t, \Delta t)\, | \, r_{\delta t}(t)< \rho \big\}_t
       & \mbox{if } \rho< 0 
  \end{cases}.
\end{align}
\end{subequations}

A comparison of $C_0(+|\rho|,\delta t, \Delta t)$ and
$C_0(-|\rho|,\delta t, \Delta t)$, should in principle be able to
reveal potential difference in the level of stock-stock correlations
during periods of raising and falling market conditions. If it is
found that $C_0(\rho,\delta t, \Delta t)$ is symmetric with respect to
the sign of $\rho$, the stock-stock correlations do not depend (very
much) on the direction of the market. On the other hand, if an
asymmetry is observed in $C_0(\pm|\rho|,\delta t, \Delta t)$ for a
given $|\rho|$, this clearly indicates that stock-stock correlations
are dependent on market direction.  Such results, being interesting in
it own right, can practically be used in risk and portfolio
management. Moreover, such results can be used as valuable input for
developing more sophisticated portfolio theories aiming at designing
the optimal portfolio. The weights of securities in an optimal
portfolio as modeled by Markowitz~\cite{16} depend on the correlations
and covariance matrices between the returns of those securities and
these correlations assume a uniform attitude towards risk.  Our
results suggest that these correlation matrices should take into
account the asymmetry in the correlations for the positive and
negative returns and, therefore, are consistent with behavioral
portfolio theory~\cite{17} that suggests different attitudes towards
risk in different domains for the same investor.

\smallskip
Given that subtle nature of the correlations that we here are trying
to detect, we will introduce an additional time average --- now to be
performed over the time scale $\delta t$ that all previously
introduced correlation functions depend. The {\em averaged conditional
  market component correlation function} is defined as
\begin{align}
   C(\rho,\Delta t) 
   &= \Big<  \big\{ C_0(\rho, \delta t, \Delta t) \big\}_{\delta t= \delta t_1}^{\delta t_2} \Big>,
  \label{eq:6}
\end{align}
where $\delta t_1$ and $\delta t_2>\delta t_1$ are time-scales over
which stock-stock correlations are relevant (given the type of data
being analyzed).  The average over $\delta t$ in Eq.~(\ref{eq:6}) is
performed {\em only} with the purpose of improving the statistics. For
stock indices containing a large number of stocks ({\it e.g.} SP500
and NASDAQ), this average may not be needed. However, for the DJIA
that currently contains only $30$ stocks, this average is of
advantage.

\bigskip

The needed formalism is by now introduced, and we are ready to use it
for the empirical analysis. Here we are focusing on the DJIA, as
mentioned previously, and the data to be analyzed were obtained from
Yahoo Finance~\cite{18}. The data set consists of daily closing prices
of the $30$ DJIA stocks as well as the DJIA index itself. It covers an
$18$~years period from May $1991$ to September $2008$. Note that this
period includes the development of the dot-com bubble in the late
$1990$’s and its subsequent burst in $2000$, the $1997$ mini-crash (as
a consequence of the Asian financial crisis of $1997$), the collapse
of the Long-Term Capital Management (as a consequence of the Russian
financial crisis of $1998$), the early $2000$’s recession as well as
the worldwide economic-financial crisis of $2007$--$2008$.

\begin{figure}[t]
  \centering 
  \includegraphics*[width=0.5\columnwidth]{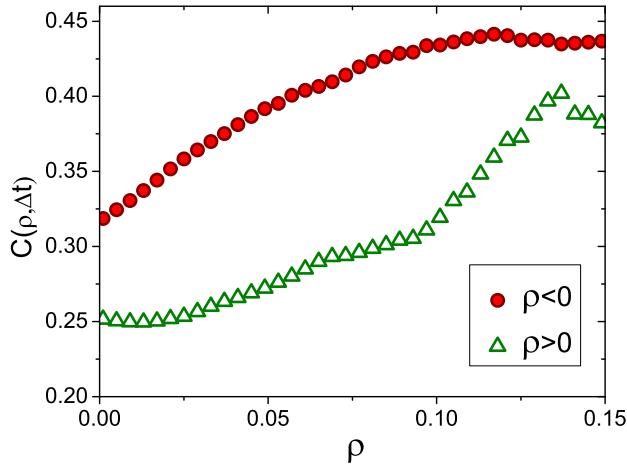} 
  \caption{The average conditional market component correlations,
    $C(\rho,\Delta t)$, between the stock components for
    various return rates, $\rho$, of the DJIA stock index. Open green
    triangles correspond the positive return levels ($\rho>0$), while
    filled red circles signifies negative return levels ($\rho<0$).
    The stronger correlation in case of negative returns are readily
    visibly from this plot. In obtaining these results,
      it was assumed that $\delta t_1=10\mbox{\,day}$, $\delta
      t_2=35\mbox{\,day}$, and $\Delta t=1\mbox{\,day}$. For values of
      $|\rho|$ larger than about $0.15$, the statistics became
      poor. This was in particular the case for positive values of $\rho$. 
    }  
  \label{Fig:3}
\end{figure}

With these data and the formalism presented previously, the averaged
conditional market component correlation function, $C(\rho,\Delta t)$,
can be calculated. It is presented in Fig.~\ref{Fig:3} for a range of
positive and negative return levels, $\pm|\rho|$, where it has been
assumed that $\Delta t=1\mbox{\,day}$, $\delta t_1=10\mbox{\,day}$ and
$\delta t_2=35\mbox{\,day}$. Figure~\ref{Fig:3} shows a pronounced
{\em asymmetry} between positive and negative (index) return levels,
$\pm|\rho|$. The stock-stock correlations, as given by $C(\rho,\Delta
t)$, are systematically stronger whenever the market is
dropping~($\rho<0$) than when it is raising~($\rho>0$). This is found
to be the case for the whole range of considered levels of return
$|\rho|$. It also worth noting that in the limit $| \rho | \rightarrow
0$ there is a substantial difference between the conditional market
component correlation for the positive and negative returns:
$\lim_{|\rho|\rightarrow 0^+} [C(-|\rho|,\Delta t) - C(|\rho|,\Delta
t)] \approx 0.07 = 7\%$.  For the largest positive levels shown, it is
noted that the statistical quality of the data is seen to become poor.

Hence, the empirical results of Fig.~\ref{Fig:3} support the primary
assumption underlying the fear-factor model~\cite{13}; stocks are on
average more strongly correlated (or synchronized) among themselves
during falling than raising market conditions.

\bigskip

The effects that we are studying here are rather subtle features, and
several averaging had to be considered in order to identify it.
Hence, it is important to have confidence in the results, and to make
sure that they are not artifacts of the averaging procedure.
Moreover, one also has to prove that the obtained difference is a
general feature of the stock market and is not due to one (or a few)
special events where {\it e.g.} the market crashes. To address these
issues, additional analysis is required:

\smallskip

Firstly, we revisited the averaging procedure over stock-stock pairs
used in defining Eq.~(\ref{eq:4}). The aim was to show that the
difference obtained in the measured correlations between the stocks
for positive and negative levels of index return was indeed present
for the majority of the stock pairs. For this purpose, for each pair
of stocks $(x,y)$ of the index, the average $C_{(x,y)}(\rho,\Delta t)=
\left< \left\{ {\mathbb C}_{(x,y)}(\rho,\delta t, \Delta t)
  \right\}_{\delta t=\delta t_1}^{\delta t_2} \right>$ was considered,
where the conditional stock-stock correlation function, ${\mathbb
  C}_{(x,y)}(\rho,\delta t, \Delta t)$, is defined from
$S_{(x,y)}(t,\delta t, \Delta t)$ in a completely analogous way to how
$C_0(\rho,\delta t, \Delta t)$ was obtained from $S_0(t,\delta t,
\Delta t)$ in Eq.~(\ref{eq:5b}).

\begin{figure}[t]
  \begin{center} 
  \subfigure {\includegraphics[width=0.30\columnwidth,height=0.30\columnwidth]{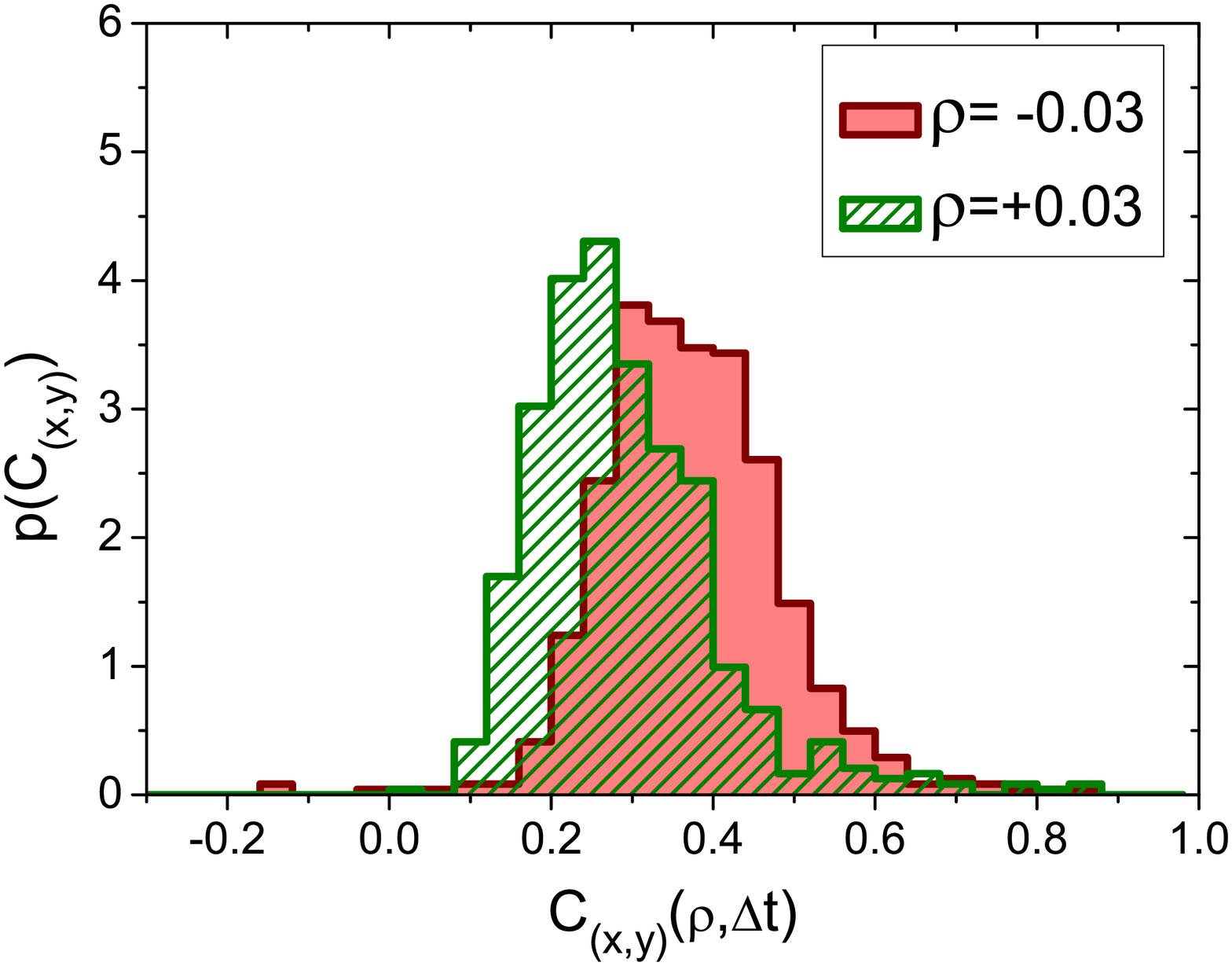}} 
  \subfigure {\includegraphics[width=0.30\columnwidth,height=0.30\columnwidth]{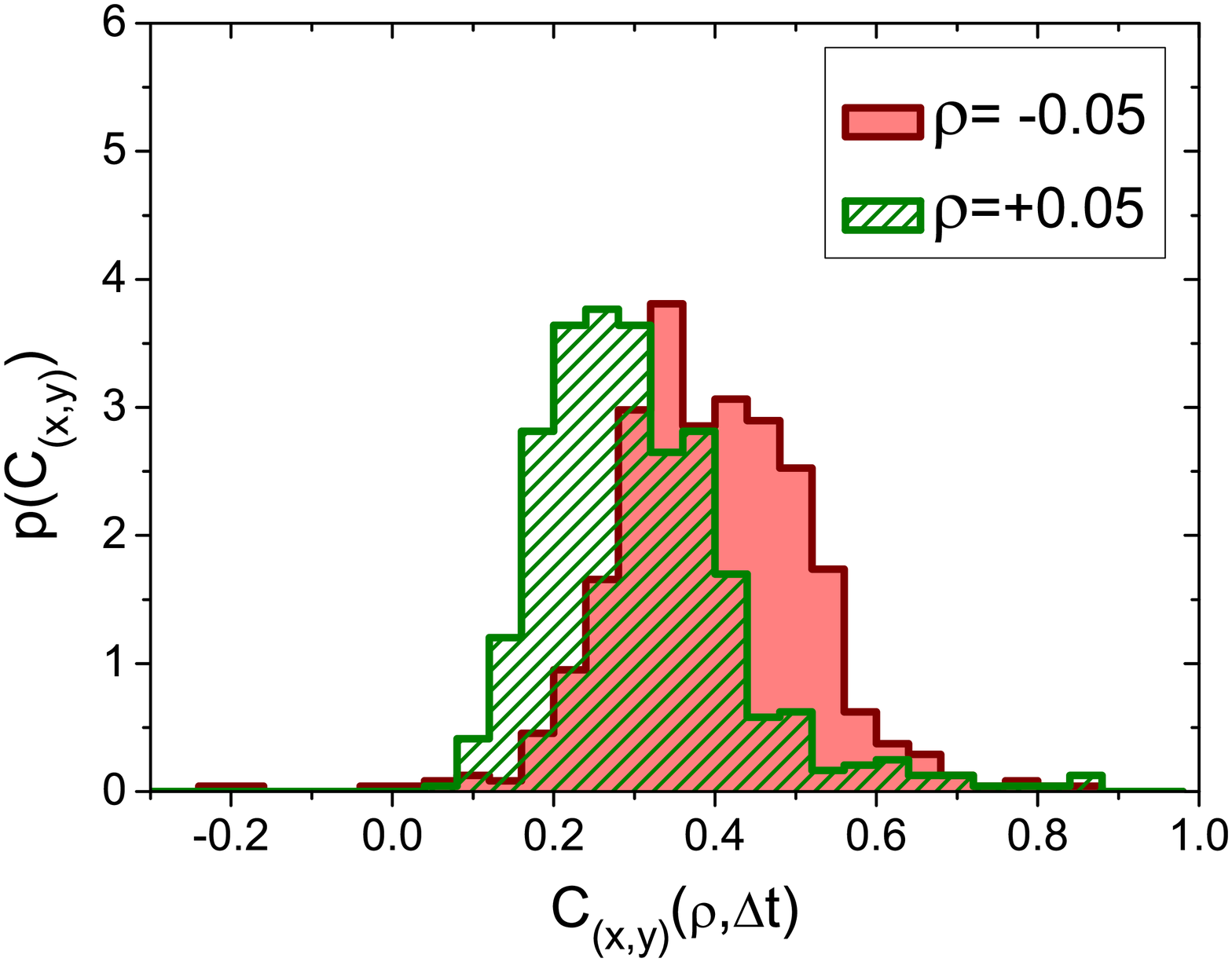}} 
  \subfigure {\includegraphics[width=0.30\columnwidth,height=0.30\columnwidth]{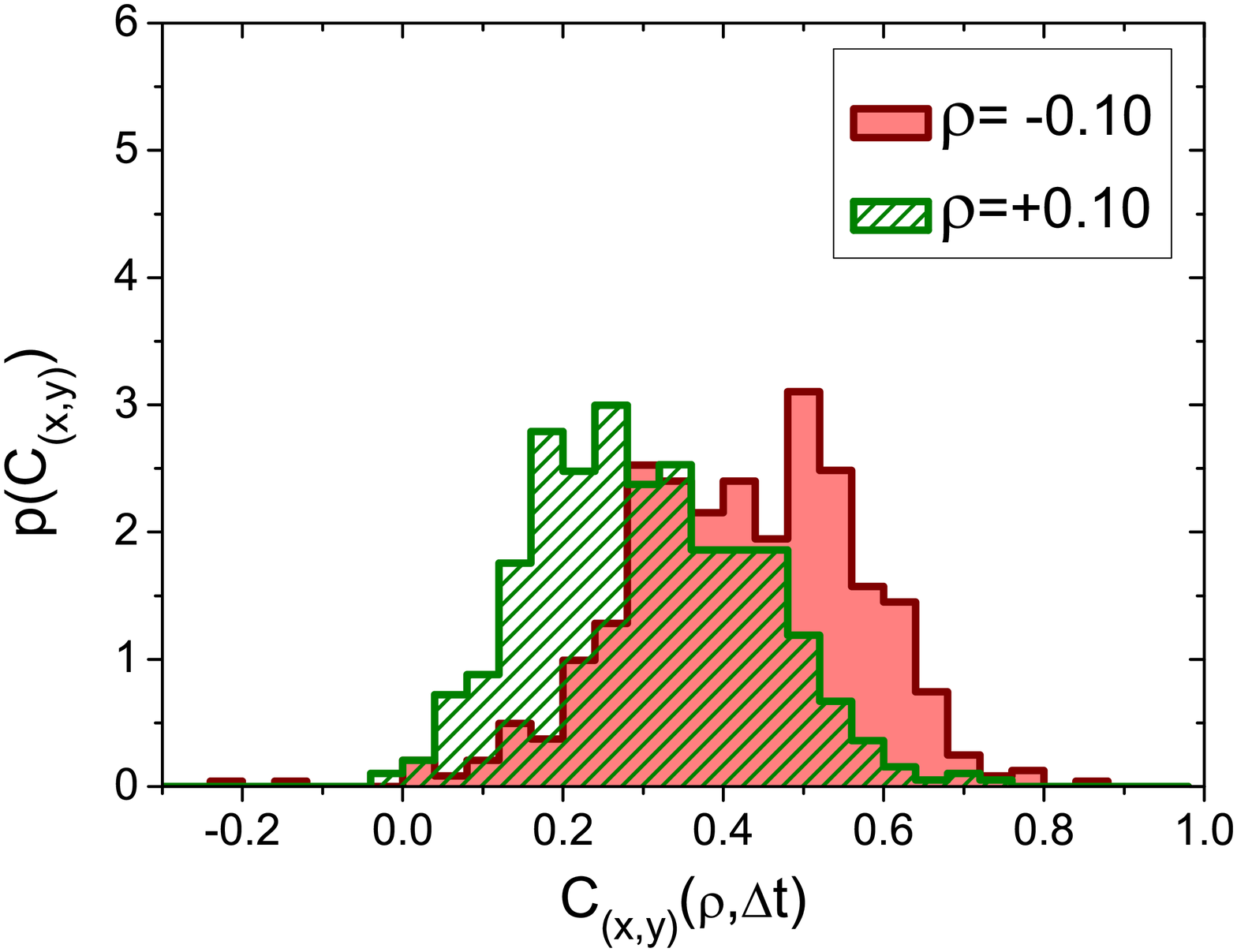}}
  \end{center}
  \caption{The distribution, $p(C_{(x,y)})$, of the correlation
    function $C_{(x,y)}(\rho,\Delta t)$ based on all
    possible company pairs $(x,y)$ within the DJIA stock index ($x
    \neq y$).  The dashed areas correspond to $\rho>0$, while the
    shaded areas refer to $\rho<0$.  The distributions are given for
    various values of the return level $\rho$ as indicated in each
    panel ($\Delta t=1\mbox{ day}$ in all cases).  
   }

  \label{Fig:4}
\end{figure}

\begin{figure}[t]
  \begin{center}
  \subfigure {\includegraphics[width=0.30\columnwidth,height=0.30\columnwidth]{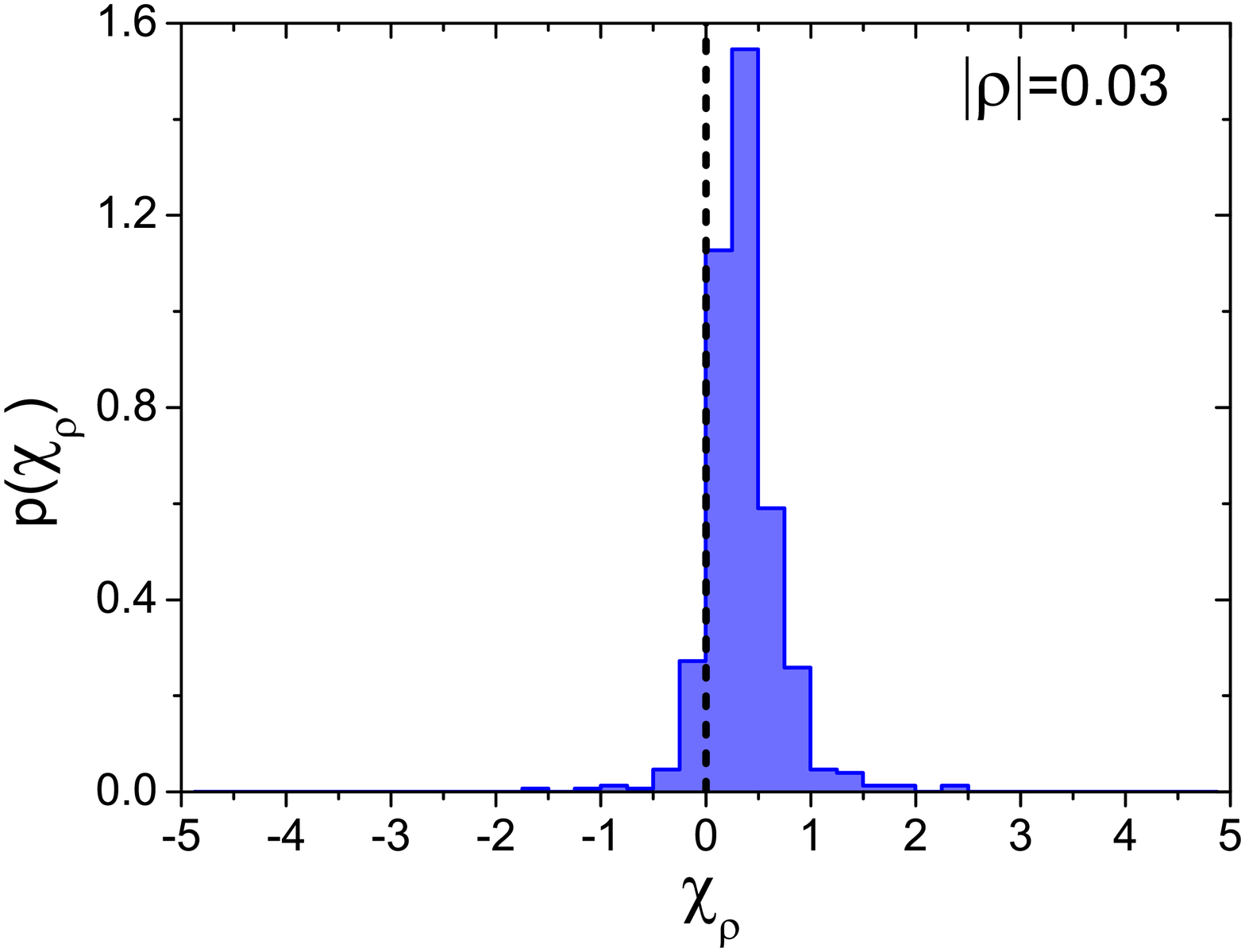}}
  \subfigure {\includegraphics[width=0.30\columnwidth,height=0.30\columnwidth]{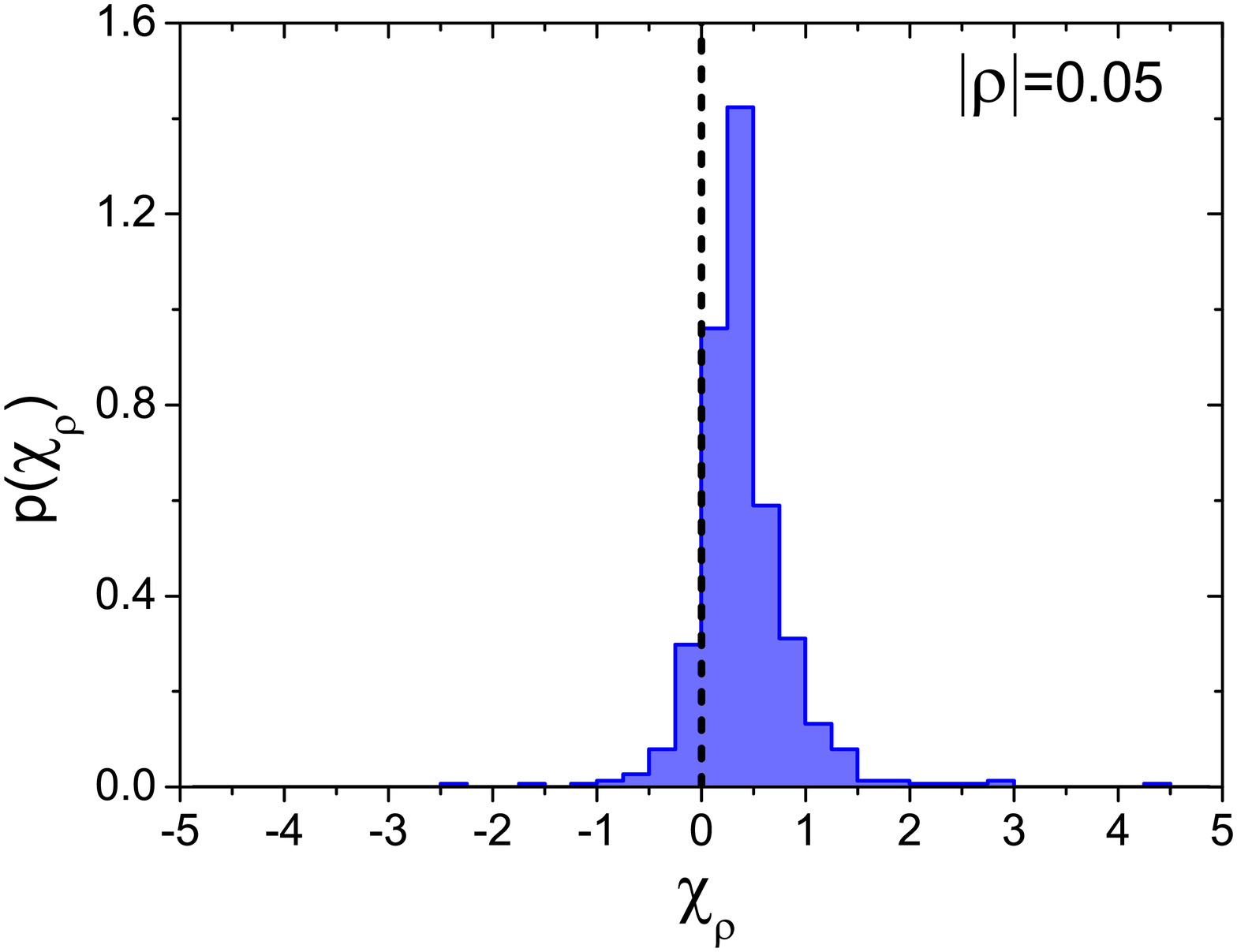}} 
  \subfigure {\includegraphics[width=0.30\columnwidth,height=0.30\columnwidth]{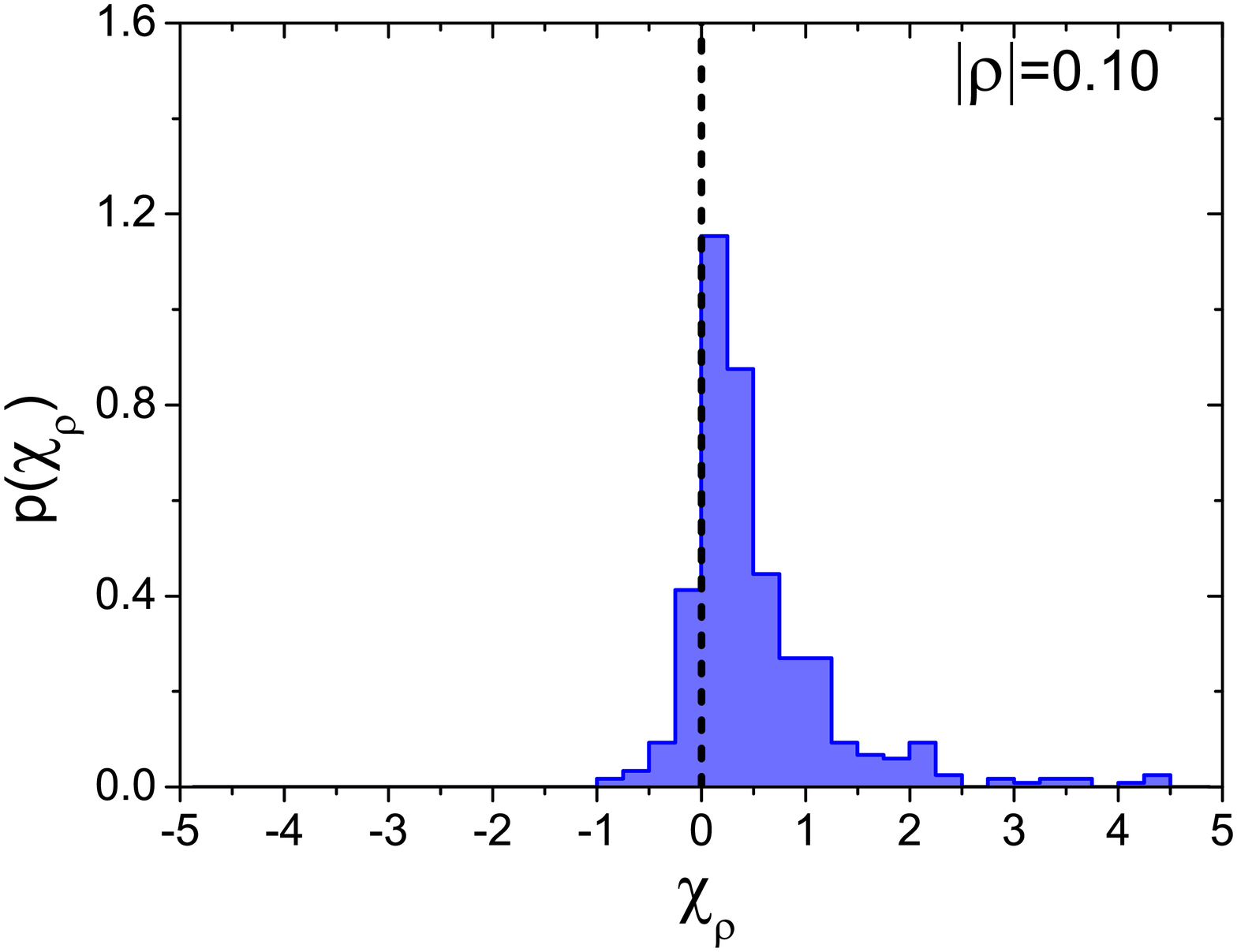}}
  \end{center}
  \caption{Distribution of the quantity 
   $\chi_\rho=[C_{(x,y)}(-|\rho|,\Delta t)-C_{(x,y)}(|\rho|,\Delta t)]/| C_{(x,y)}(|\rho|,\Delta t) | $  
   obtained on the basis of the DJIA stock index for different
   return levels, $\rho$, as indicated in the figures.
   We recall that in the case no asymmetry, the distribution,
   $p(\chi_\rho)$, will be symmetric around $\chi=0$ (vertical dashed
   lines).
 }
  \label{Fig:4b}
\end{figure}

The distributions of the conditional stock-stock correlation function,
$C_{(x,y)}(\rho,\Delta t)$, including all possible stocks pair ($x\neq
y$) of the DJIA, is presented in Figs.~\ref{Fig:4} for some
representative levels of index return $|\rho|=0.03$, $0.05$ and
$0.10$. The results of Figs.~\ref{Fig:4} indicate that the stock-stock
correlations for a negative index return levels, $-|\rho|$, plotted
with green shades is for the majority of the stock pairs stronger than
the stock-stock correlations for the corresponding positive level, and
this observation applies equally for all the index return levels
considered. An alternative way for illustrating this difference is to
plot the distribution of the relative difference
$\chi_\rho=[C_{(x,y)}(-|\rho|,\Delta t)-C_{(x,y)}(|\rho|,\Delta t)]/|
C_{(x,y)}(|\rho|,\Delta t) |$ (Figs.~\ref{Fig:4b}). The clear
asymmetry of this distribution respective to $0$ is an indication that
the stock-stock correlations for a negative index return level is in
general stronger than the stock-stock correlations for the
corresponding positive level.

The indications obtained from Figs.~\ref{Fig:4} and Figs.~\ref{Fig:4b}
that the conditional stock-stock correlations are stronger for
negative index return levels can also be confirmed more quantitatively
by a statistical test. More precisely, we want to see what is the
chance that two random samples from the same distribution would yield
the observed difference in the mean. A Wilcoxon-type non-parametric
$z$-test~\cite{19} was performed and the results of the test are
presented in Table~\ref{Tab:1}. The negative value of $z$ suggests
that the stock-stock correlations for the negative change in the index
are indeed bigger than those for the positive changes. The value of
$p$ is the probability that finite samples from the same ensemble
would yield the hypothesized differences in the mean. The parameter
$p$ is thus a measure of the significance level, smaller values
correspond to higher significance for the obtained differences in the
mean.  The results presented in Table~\ref{Tab:1} show that the
difference in conditional stock-stock correlations is indeed
significant.

\begin{table}[tbh]
\begin{center}
\begin{tabular}{c @{\hspace{7mm}}c @{\hspace{7mm}} c}
  \hline
  \hline
  $|\rho|$        & $z$        & $p$  \\
  \hline 
  0.03                & $-18.87$     &  $2.0\cdot 10^{-79}$  \\
  0.05                & $-18.16$     &  $9.1\cdot 10^{-74}$  \\
  0.10                & $-10.85$     &  $1.8\cdot 10^{-27}$ \\
  \hline
  \hline
\end{tabular}
\end{center}

\caption{
  Results of the Wilcoxon non-parametric $z$-test for difference in
  conditional stock-stock correlations ($C_{(x,y)}(\rho,\Delta t)$).  The negative
  $z$-values suggest that $C_{(x,y)}(-|\rho|,\Delta
  t)>C_{(x,y)}(|\rho|,\Delta t)$.
  The value of $p$ is the probability that a finite sample
  taken from the same ensemble would yield the hypothesized difference
  in the mean.}  
\label{Tab:1}
\end{table}

\smallskip
Secondly, we wanted to make sure that the observed asymmetry in
$C_0(\rho,\delta t, \Delta t)= \langle \left\{ {\mathbb
    C}(\rho,t,\delta t,\Delta t)\right\}_t\rangle$ [Eq.~(\ref{eq:5})]
was not caused by a few isolated events --- like large market drops
--- but instead represented a feature of the market that was present
at (more-or-less) all times. For this purpose, we went back and
studied more carefully the time-dependent conditional market
correlation function ${\mathbb C}(\rho,t,\delta t,\Delta t)$ (before
the time average).  More precisely, in order to improve the
statistics, the following average was computed $\left< \left\{
    {\mathbb C}(\rho,t,\delta t,\Delta t)\right\}_{\delta t = \delta
    t_1}^{\delta t_2} \right> \equiv C_t(\rho, \Delta t)$. For fixed
values of the index return level $|\rho|$, and the time windows
$\delta t_1=10$ days, $\delta t_2=35$ days and $\Delta t=1$ day, we
compared the two distributions $p[C_t(+|\rho|, \Delta t)]$ and
$p[C_t(-|\rho|, \Delta t)]$. An asymmetry in $C_0(\rho,\delta t,
\Delta t)$ being caused by a few isolated events in $C_t(\rho, \Delta
t)$, will produce almost identical distributions for the two cases
$\pm|\rho|$ that only differ by some infrequent ``outliers'' that are
large enough to move the mean. On the other hand, a more systematic
difference in $C_t(\rho, \Delta t)$ for $+|\rho|$ and $-|\rho|$ will
produce distinctly differences between the $p[C_t(+|\rho|, \Delta t)]$
and $p[C_t(-|\rho|, \Delta t)]$ distributions.

\begin{figure}[t]
  \begin{center}
  \subfigure {\includegraphics[width=0.30\columnwidth,height=0.30\columnwidth]{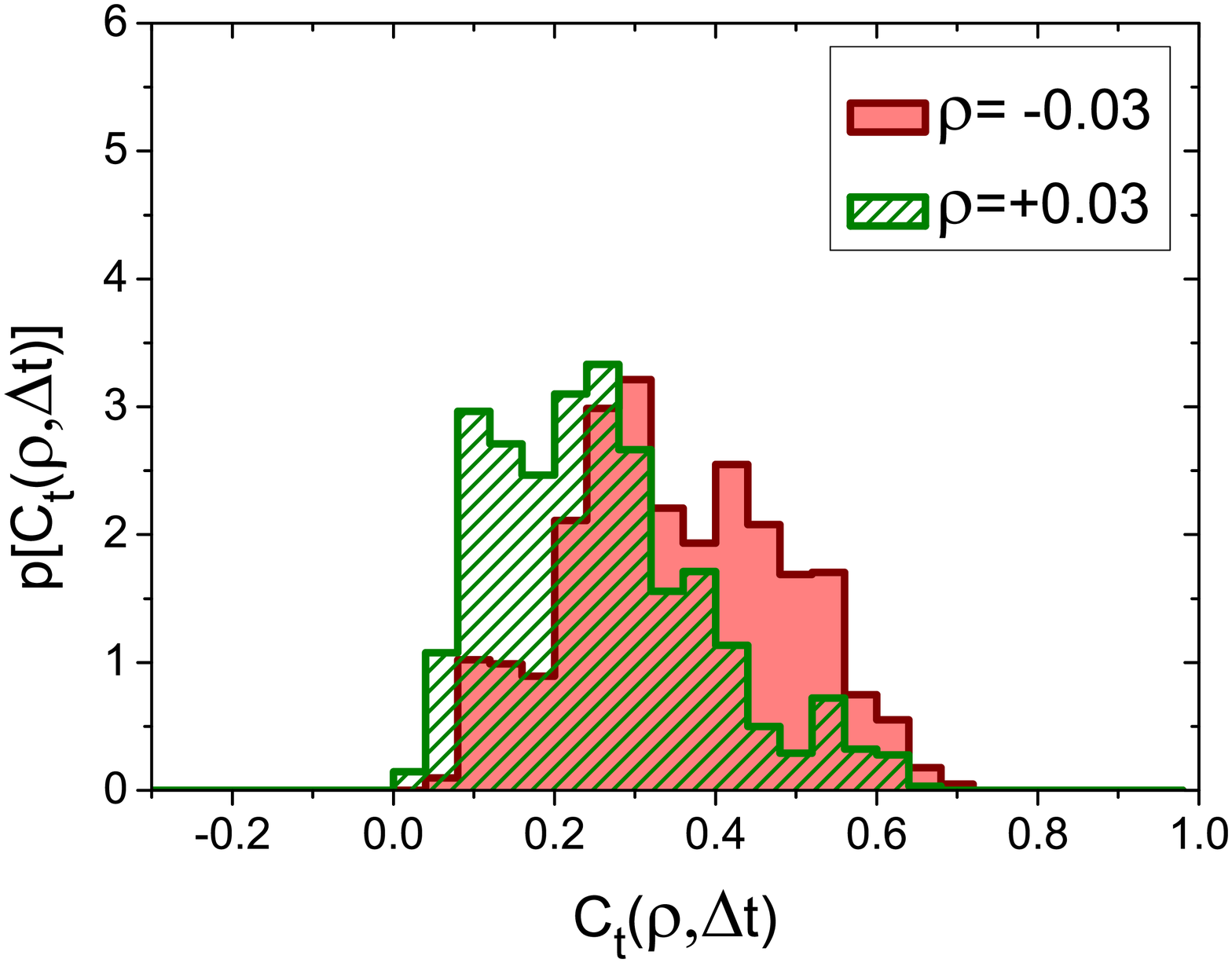}}
  \subfigure {\includegraphics[width=0.30\columnwidth,height=0.30\columnwidth]{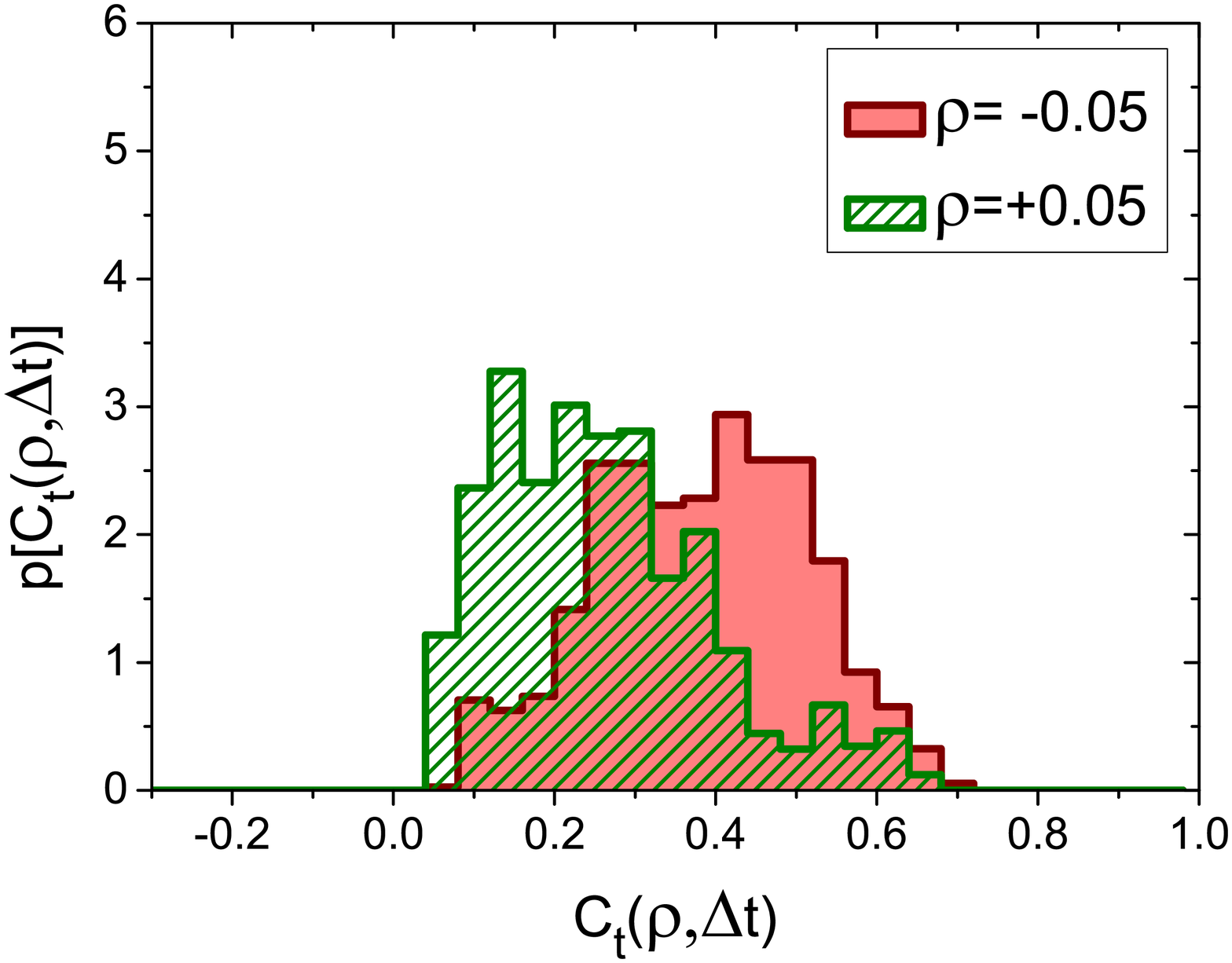}}
  \subfigure {\includegraphics[width=0.30\columnwidth,height=0.30\columnwidth]{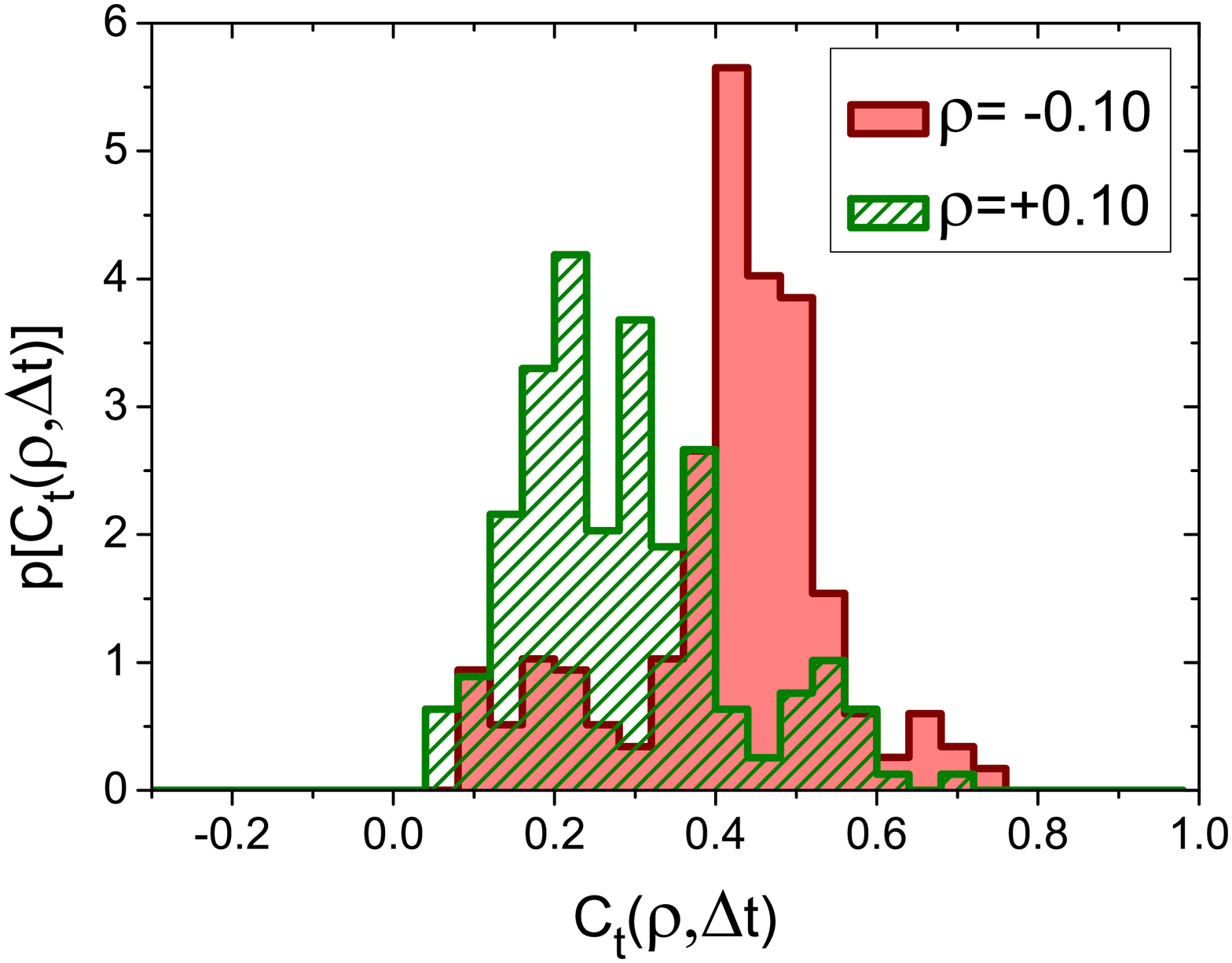}}
  \end{center}
  \caption{Distribution of the correlations $C_t(\rho, \Delta t) \equiv \left< \left\{
        {\mathbb C}(\rho,t,\delta t,\Delta t)\right\}_{\delta t =
        \delta t_1}^{\delta t_2} \right>$ for different return levels,
    $\pm |\rho|$. Here the green shaded areas correspond to
      $\rho>0$ while red is used to indicate $\rho<0$. In obtaining
      these results it was assumed that $\delta t_1=10\mbox{\,day}$,
      $\delta t_2=35\mbox{\,day}$, and $\Delta t=1\mbox{\,day}$.
}
  \label{Fig:5}
\end{figure}

In Figs.~\ref{Fig:5} we present the empirical distributions
$p[C_t(\rho. \Delta)]$ of the DJIA for some typical positive and
negative values of the index return level. These empirical results
point towards the two distributions $p[C_t(+|\rho|, \Delta t)]$ and
$p[C_t(-|\rho|, \Delta t)]$ being different.  To quantitatively show
that they differ significantly, again a non-parametric Wilcoxon
significance test was performed. However, in order to conduct this
test it is necessary to have the same number of data points in the
histograms for positive and negative index return values. Since the
$C_t(\rho, \Delta)$-data did not had this property we had to ensure
this condition.  We first identified the set with the smallest number
of elements (usually this was the set corresponding the negative
returns), and then from the other set, the same number of elements
were randomly selected. Here, our assumption was that the random
selection will not alter the normalized distribution.  Results
obtained by this procedure for the same values of $|\rho|$ used to
produce Figs.~\ref{Fig:5} are given in Table~\ref{Tab:2}.  The
extremely small values obtained for $p$ suggest, as pointed out
previously, that the difference between the two distributions,
$p[C_t(+|\rho|, \Delta t)]$ and $p[C_t(-|\rho|, \Delta t)]$, is indeed
significant also for this averaging step.

\begin{table}[tbh]
\begin{center}
\begin{tabular}{c @{\hspace{7mm}}c @{\hspace{7mm}} c}
  \hline
  \hline
  $|\rho|$        & $z$        & $p$ \\
  \hline
  0.03                & $-33.99$     &  $1.1\cdot 10^{-79}$  \\
  0.05                & $-16.62$     &  $4.3\cdot 10^{-62}$  \\
  0.10                & $-8.0$     &  $1.0\cdot 10^{-16}$ \\
  
  \hline
  \hline
\end{tabular}
\end{center}

\caption{ Results of the Wilcoxon non-parametric $z$ test for the
  difference in the mean of the distributions 
  $p[C_t(+|\rho|, \Delta t)]$ and $p[C_t(-|\rho|, \Delta t)]$,
  presented in Figs.~\protect\ref{Fig:5}}
\label{Tab:2}
\end{table}

\begin{figure}[t]
  \begin{center}
  \subfigure {\includegraphics[width=0.30\columnwidth,height=0.30\columnwidth]{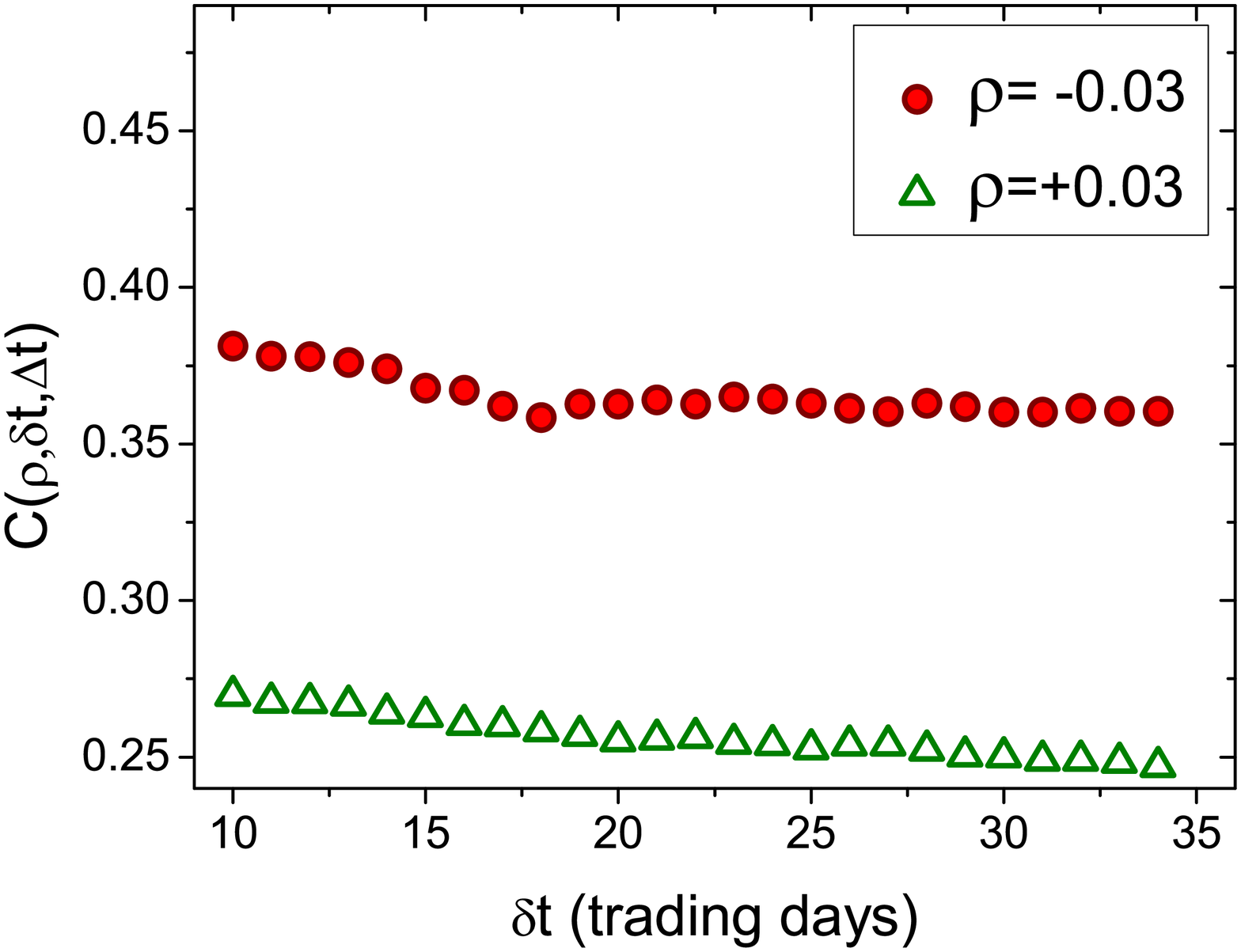}}
  \subfigure {\includegraphics[width=0.30\columnwidth,height=0.30\columnwidth]{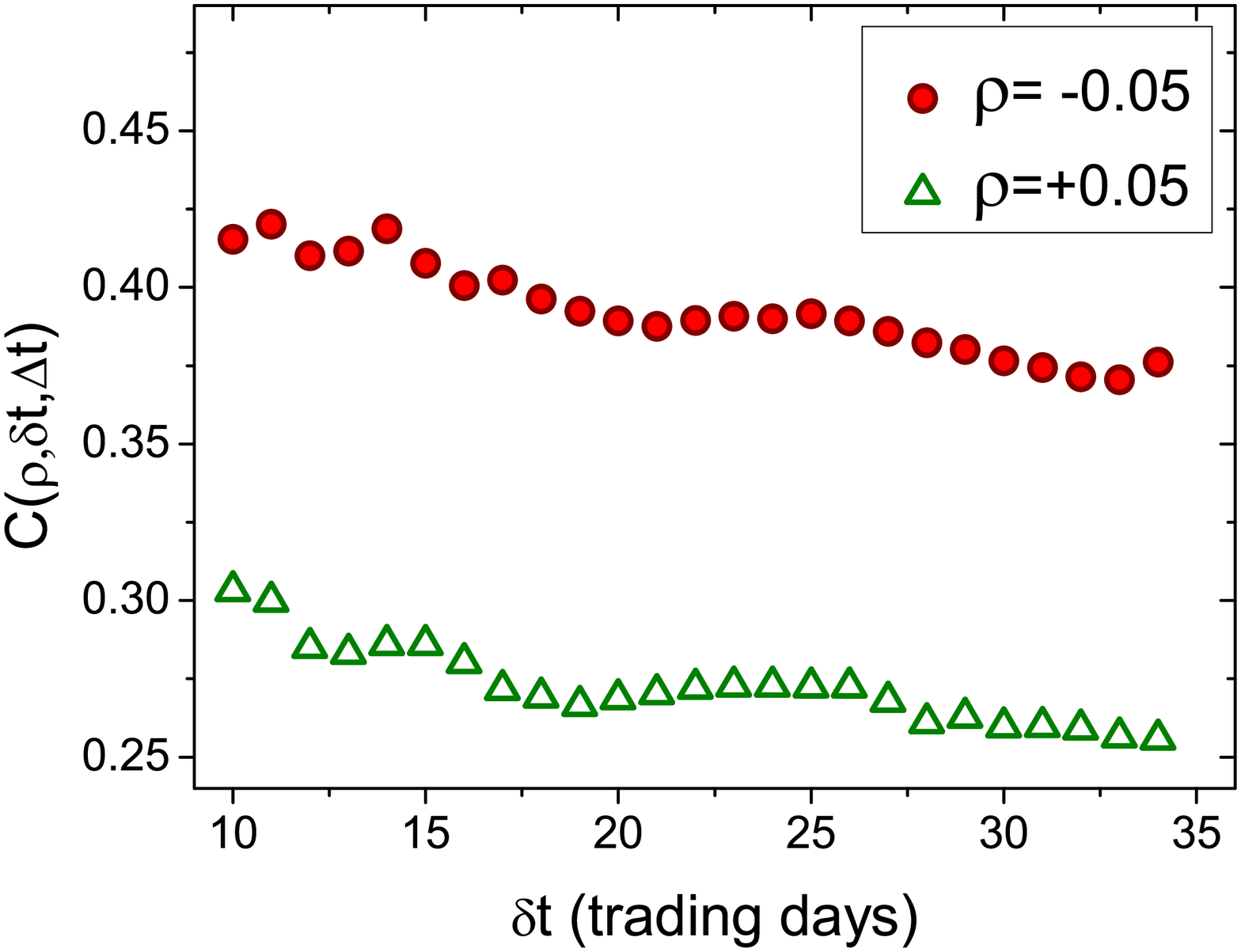}}
  \subfigure {\includegraphics[width=0.30\columnwidth,height=0.30\columnwidth]{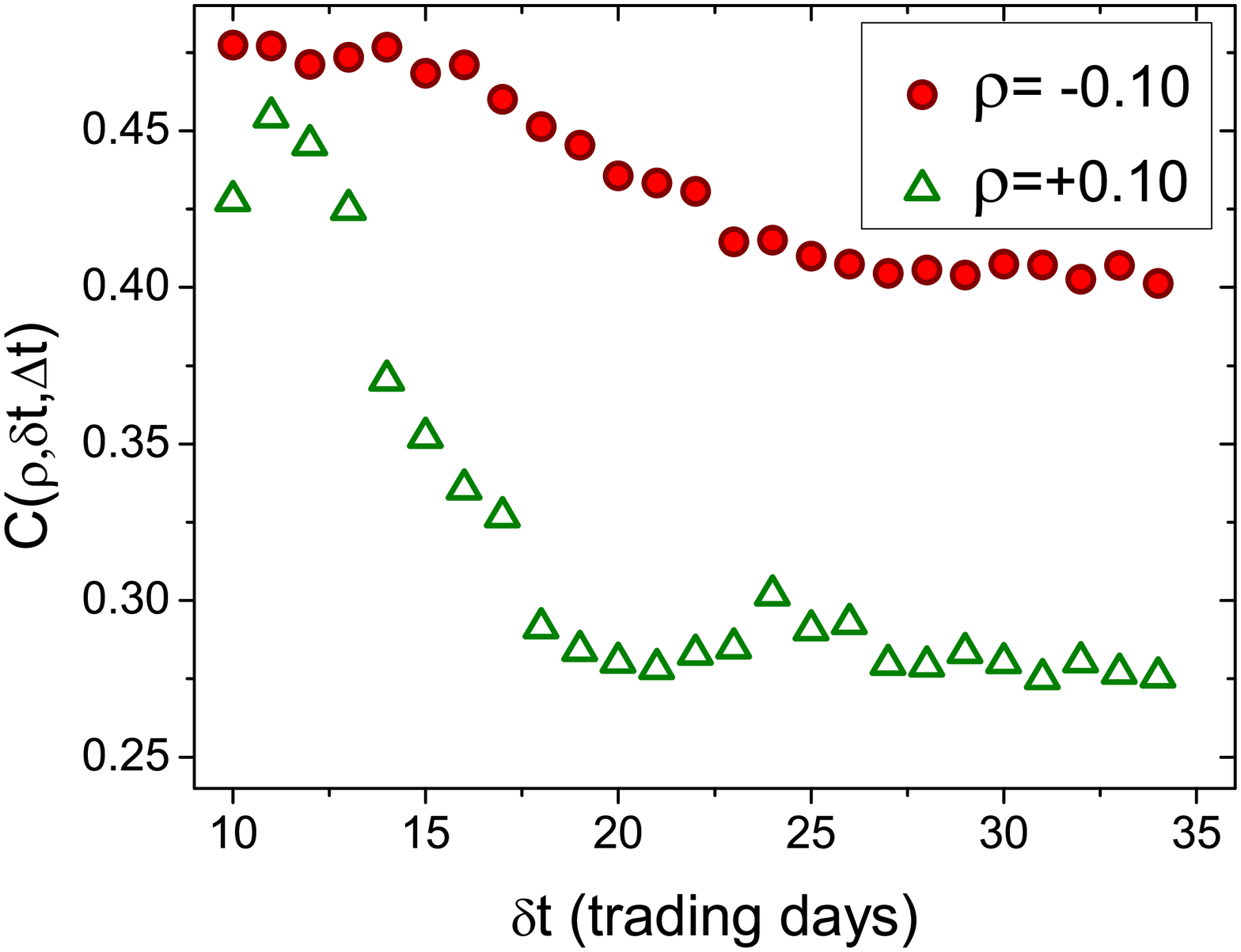}}
  \end{center}
  \caption{Conditional market correlation, $C_0(\rho,\delta t, \Delta
    t)$, as a function of the time-window $\delta t$ (with
    $\Delta t=1\mbox{\,day}$) for different values of the return level
    $\pm|\rho|$. Green open triangles correspond to
    $\rho>0$ while filled red circles refer to $\rho<0$.
        }  
\label{Fig:6}
\end{figure}

\smallskip
Thirdly, and finally, we address the level of conditional market
correlation ($C_0(\rho,\delta t, \Delta t)$) as a function of the size
of the time-window $\delta t$ for $\pm|\rho|$ (and $\Delta
t=1\mbox{\,day}$). The empirical results of this kind are depicted on
Figs.~\ref{Fig:6}. One observes that systematically, and independent
of $\delta t$ and $\rho$ (at lest for the values we have considered),
one finds that the conditional market correlations are the higher for
negative index return levels ($-|\rho|$) as compared to the
corresponding positive ones ($+|\rho|$); {\it i.e.}
$C_0(-|\rho|,\delta t, \Delta t) > C_0(+|\rho|,\delta t, \Delta t)$.
This suggests that the sign of the difference does not depend on the
values considered for $\delta t_1$ and $\delta t_2$, used in
performing the average over $\delta t$.

\smallskip 
In conclusion, we have conducted a set of statistical investigations
on the DJIA and its constituting stocks, which confirm that during
falling markets, the stock-stock correlations are stronger than during
market raises (gain-loss asymmetry phenomenon). This has been possible 
to measure empirically due to the design of a robust statistical measure --- 
the conditional market correlation function ($C_0(\rho,\delta t, \Delta t)$).

In particular, we have performed statistical tests that show that the
observed asymmetry in the empirical conditional (market) correlation
function is indeed significant, and not an artifact of the considered
averaging procedure since it is clearly present in each averaging
step. This empirical result gives confidence in the fear-factor
hypothesis, which explains successfully the gain-loss asymmetry
observed in the major stock indices.

From the perspective of finance, we note that a relatively small
segment of the financial literature examines models which have the
potential to describe, explain and possibly forecast the phenomena
which lead to stock market bubbles and their subsequent
crashes~\cite{Shiller}.  The more technical and quantitative
approaches either follow the general equilibrium models of
macroeconomics~\cite{Blanchard} or the game-theoretical
methodology~\cite{Brunnermeier}.

The latter approaches try to model mathematically (many times using
toy models) the interactions between agents and their expectations
about each other's behavior and the market average. Many times market
micro-structure plays a significant role in these models: the
so-called frictions (the different taxes and transaction costs,
liquidity constraints and other limits to arbitrage) are the factors
that produce market crashes. The role of portfolio insurance (selling
short the stock index futures, \cite{Brennan-Schwartz}) in crashes is
also strongly debated. However, the complex relationship between the
micro-structure factors, market sentiment, herding of investors and
stock market crashes is still poorly understood.

In such view our results can have important consequences in
theoretical and practical aspects of portfolio management and also in
risk management of investment banks, investment funds, other financial
institutions as well as regulators and decision makers concerned with
the spillover of stock market crashes into the real economy.  As it
was pointed out earlier, the standard, mean-variance based portfolio
theory views risks as symmetric measures (variance, covariance, {\it
  etc}.) assuming the stability of these risks as well as their
symmetry in case of positive and negative returns. Investment banks,
insurance companies and other financial institutions widely use for
risk management software based on the methodology of VaR (Value at
Risk), a measure of worst-case scenario losses that is intensively
questioned since today’s financial crisis began.  VaR models the case
of symmetrical risks, relying in most cases on past distributions
(especially on the normal distribution). Over-reliance on VaR lead the
risk managers to the following mistakes: ({\it i}) It leads to the
opening and maintaining of risky and overly leveraged positions; ({\it
  ii}) It focused on the manageable risks with probabilities close to
the center of the probability distributions and it lost track of the
extreme events from the tails of the distributions; ({\it iii})
Utilization of VaR leads to a false sense of security among risk
managers.  We believe that new measures must be considered instead of
VaR.  These measures should also take into account the fear-factor
which produces bigger systematic risks in cases of stock market
crashes than during market booms. As a follow up, it will be worthy to
study whether a growing distance between the negative and positive
correlations is a sign of diminishing investor confidence in the
periods before a market crash.

\acknowledgments
One of the authors (I.S.) is grateful for constructive discussions
with J.D. Farmer and M.H. Jensen on the matter of this Letter.  This
research was supported in part by a KMEI-Interdisciplinary Computer
Modelling research grant and the work of E.B. is supported by a
Scientific Excellence Bursary from BBU.  I.S. acknowledges the support
from the European Union COST Actions P10 (``Physics of Risk'') and
MP0801 (``Physics of Competition and Conflicts'').





\begin{thebibliography}{99}


\bibitem{1} N.F. Johnson, P. Jefferies and P.M. Hui, {\sl Financial
    Market Complexity} (Oxford University Press, 2003).

\bibitem{2} 
  J.P. Bouchaud  and M. Potters,
  {\sl Theory of Financial Risks: From Statistical Physics to Risk Management} 
  (Cambridge University Press, Cambridge, 2000).

\bibitem{3}  R.N. Mantegna and H.E. Stanley,
  {\sl An  Introduction to Econophysics: Correlations and Complexity
    in Finance}
  (Cambridge University Press, Cambridge, 2000).


\bibitem{Bodie}
Z. Bodie, A. Kane, A. J. Marcus, {\sl Investments, 8th edition}, 
(McGraw Hill New York, 2008)


\bibitem{Shiller}
R.J. Shiller, {\sl Irrational exuberance}, 2nd ed. (Princeton
University Press, 2005).



\bibitem{Sornette} D. Sornette, {\sl Why Stock Markets Crash: Critical
    Events in Complex Financial Systems} (Princeton University Press,
  2002).


\bibitem{4} I. Simonsen, M.H. Jensen and  A. Johansen,
  Eur. Phys. J.  {\bf 27}, 583--586 (2002).

\bibitem{5} 
  M.H. Jensen, A. Johansen and I. Simonsen, Physica A {\bf 234}, 338--343 (2003).

\bibitem{6} 
  M.H. Jensen, A. Johansen, F. Petroni and I. Simonsen,
  Physica A {\bf 340}, 678--684 (2004).

\bibitem{7} 
  M.H. Jensen, A. Johansen and I. Simonsen,
  Int. J. Mod. Phys. B  {\bf 17}, 4003--4012 (2003).

\bibitem{8} 
  M.H. Jensen, Phys. Rev. Lett. {\bf 83}, 76--79 (1999).


\bibitem{8-2}
  Wei-Xing Zhou, D. Sornette, and Wei-Kang Yuan
  Physica D {\bf 214}, 55 (2006).

\bibitem{9} S. Redner, {\sl A guide to first-passage processes}
  (Cambridge University Press, Cambridge, 2001).


\bibitem{10}
  Wei-Xing Zhou,  and Wei-Kang Yuan
  Physica A {\bf 353}, 433 (2005).

\bibitem{12} A. Johansen, I. Simonsen and M. H. Jensen,
  Physica A {\bf 370}, 64 (2006).

\bibitem{12A}
  I. Simonsen, P.T.H. Ahlgren, M.H. Jensen, R. Donangelo, and K. Sneppen 
  Eur. Phys. J. B {\bf 57}, 153 (2007).



\bibitem{Poland}
  M. Zaluska-Kotur, K. Karpio, and A. Or{\l}owska,
  Acta Phys. Pol. B {\bf 37}, 3187 (2006).

\bibitem{Austria_Poland}
  K. Karpio, M.A. Za{\l}uska-Kotur, and A. Or{\l}owski,         
  Physica A {\bf 375}, 599 (2007).

\bibitem{Korea}
  C.Y. Lee, J. Kim, and I. Hwang,
  J. Kor. Phys. Soc. {\bf 52}, 517 (2008).

\bibitem{Poland-2}
  M. Grudziecki, E. Gnatowska, K.  Karpio, A. Or{\l}owska,  and M. Za{\l}uska-Kotur,
  Act. Phys. Pol. A {\bf 114}, 569 (2008).


\bibitem{Saxo-2}
  J.V. Siven and J.T. Lins,
  Phys. Rev. E {\bf 80}, 057102 (2009).



\bibitem{Peter}
  P.T.H. Ahlgren, H. Dahl, M.H. Jensen, and I. Simonsen, 
  {\sl What Can Be Learned from Inverse Statistics?} 
  (Springer-Verlag, 2010) Book chapter to appear.


\bibitem{Our_Unpublished} 
  Johansen, I. Simonsen, and M.H. Jensen, Unpublished work.

  




\bibitem{13}  
  R. Donangelo, M.H. Jensen, I. Simonsen and K. Sneppen, 
  J. Stat. Mech. {\bf 2006}, L11001 (2006). 

\bibitem{Saxo-1}
  J.V. Siven, J.T. Lins, J.L. Hansen,
  J. Stat. Mech. {\bf 2009}, P02004 (2009).

\bibitem{Nelson}
  D.B. Nelson,
   Econometrica, {\bf 59}, 347 (1991).

\bibitem{Leverage}
  J.-P. Bouchaud, A. Matacz, M. Potters, 
  Phys. Rev. Lett. {\bf 87}, 228701 (2001). 

\bibitem{Peter_Leverage}
  P.T.H. Ahlgren, M.H. Jensen, I. Simonsen, R. Donangelo, and K. Sneppen 
  Physica A {\bf 383}, 1 (2007).

\bibitem{Saxo-3} 
  J.V. Siven and J.T Lins, 
  {\sl Gain/loss asymmetry in time series of individual stock prices and its relationship to the leverage effect}
  arXiv:0911.4679v2, 2009.


\bibitem{14} 
  S. Strogatz, Physica D {\bf 143}, 1 (2000).

\bibitem{15} 
  D.Kahneman and  A. Tversky, 
  Econometrica  {\bf 47}, 313--327 (1979).

\bibitem{16}
  H.M. Markowitz, 
  Journal of Finance {\bf 7}, 77--91 (1952).

\bibitem{17}
 H. Shefrin and M. Statman (2000), 
 The Journal of Financial and Quantitative Analysis, {\bf 35}, 127--151 (2000).

\bibitem{18}
  The analyzed data are freely available from the Yahoo:
  http://finance.yahoo.com. 

\bibitem{19} 
  F. Wilcoxon, Biometrics {\bf 1}, 80--83 (1945).




\bibitem{X} 
  J.V. Siven, J.T Lins, and J.L. Hansen, 
  J. Stat. Mech. {\bf  2009}, P02004 (2009)






\bibitem{Blanchard}
O.J. Blanchard, 
Econ. Lett. {\bf 3}, 387 (1979).

\bibitem{Brunnermeier}
M. Brunnermeier, 
{\sl Asset Pricing Under Asymmetric Information} (Princeton
University Press, 2001).

\bibitem{Brennan-Schwartz} 
  M.J. Brennan and E.S. Schwartz, 
  J. Bus. {\bf 62}, 455 (1989).



\end{thebibliography}
\end{document}